\documentclass[prx,epsf,showpacs,twocolumn,longbibliography,superscriptaddress,]{revtex4-1}
\usepackage[pdftex]{graphicx}
\usepackage{dcolumn}
\usepackage{bm}
\usepackage{epsfig}
\usepackage{latexsym}
\usepackage{amsmath}
\usepackage{amsfonts}
\usepackage{color}
\usepackage{array}
\usepackage{braket}
\usepackage{dsfont}
\usepackage{mathrsfs}
\usepackage{nicefrac}
\usepackage{hyperref}
\hypersetup{
    colorlinks,
    citecolor=blue,
    filecolor=red,
    linkcolor=magenta,
    urlcolor=red
}
\usepackage[dvipsnames]{xcolor}

\newcommand{\<}{\langle}

\renewcommand{\>}{\rangle}
\renewcommand{\(}{\left(}
\renewcommand{\)}{\right)}

\makeatletter
\DeclareFontFamily{OMX}{MnSymbolE}{}
\DeclareSymbolFont{MnLargeSymbols}{OMX}{MnSymbolE}{m}{n}
\SetSymbolFont{MnLargeSymbols}{bold}{OMX}{MnSymbolE}{b}{n}
\DeclareFontShape{OMX}{MnSymbolE}{m}{n}{
    <-6>  MnSymbolE5
   <6-7>  MnSymbolE6
   <7-8>  MnSymbolE7
   <8-9>  MnSymbolE8
   <9-10> MnSymbolE9
  <10-12> MnSymbolE10
  <12->   MnSymbolE12
}{}
\DeclareFontShape{OMX}{MnSymbolE}{b}{n}{
    <-6>  MnSymbolE-Bold5
   <6-7>  MnSymbolE-Bold6
   <7-8>  MnSymbolE-Bold7
   <8-9>  MnSymbolE-Bold8
   <9-10> MnSymbolE-Bold9
  <10-12> MnSymbolE-Bold10
  <12->   MnSymbolE-Bold12
}{}

\let\llangle\@undefined
\let\rrangle\@undefined
\DeclareMathDelimiter{\llangle}{\mathopen}%
                     {MnLargeSymbols}{'164}{MnLargeSymbols}{'164}
\DeclareMathDelimiter{\rrangle}{\mathclose}%
                     {MnLargeSymbols}{'171}{MnLargeSymbols}{'171}

\newcommand{\fref}[1]{Fig.~\ref{#1}}
\newcommand{\eref}[1]{Eq.\,(\ref{#1})}

\begin{document}

\title{Holographic quantum algorithms for simulating correlated spin systems}

\author{Michael Foss-Feig}
\email{michael.feig@honeywell.com}
\author{David Hayes}
\author{Joan M. Dreiling}
\author{Caroline Figgatt}
\author{John P. Gaebler}
\author{Steven A. Moses}
\author{Juan M. Pino}
\affiliation{Honeywell $|\!$ Quantum Solutions}
\author{Andrew C. Potter}
\email{acpotter@utexas.edu}
\affiliation{Honeywell $|\!$ Quantum Solutions}
\affiliation{Department of Physics, University of Texas at Austin, Austin, TX 78712, USA}

\begin{abstract}
We present a suite of ``holographic" quantum algorithms for efficient ground-state preparation and dynamical evolution of correlated spin-systems, which require far-fewer qubits than the number of spins being simulated. The algorithms exploit the equivalence between matrix-product states (MPS) and quantum channels, along with partial measurement and qubit re-use, in order to simulate a $D$-dimensional spin system using only a ($D$-1)-dimensional subset of qubits along with an ancillary qubit register whose size scales logarithmically in the amount of entanglement present in the simulated state. Ground states can either be directly prepared from a known MPS representation, or obtained via a holographic variational quantum eigensolver (holoVQE). Dynamics of MPS under local Hamiltonians for time $t$ can also be simulated with an additional (multiplicative) ${\rm poly}(t)$ overhead in qubit resources.  These techniques open the door to efficient quantum simulation of MPS with exponentially large bond-dimension, including ground-states of 2D and 3D systems, or thermalizing dynamics with rapid entanglement growth. As a demonstration of the potential resource savings, we implement a holoVQE simulation of the antiferromagnetic Heisenberg chain on a trapped-ion quantum computer, achieving within $10(3)\%$ of the exact ground-state energy of an infinite chain using only a pair of qubits.

\end{abstract}
\maketitle

\section{Introduction}
One of the most promising near-term applications of quantum computers is the simulation of correlated quantum systems in which entanglement plays a crucial role, for which accurate classical simulations are often intractable.  Examples include predicting low-temperature properties of correlated materials \cite{bauer2016hybrid}, calculating reaction rates or photoabsorption spectra of large molecules \cite{cao2019quantum}, and simulating lattice gauge theories of particle physics \cite{banuls2019simulating}.  While these applications have generated considerable excitement, it is far from clear how large and how accurate a quantum computer will need to be in order to address classically hard questions of practical scientific and technological relevance.  Many problems of interest require extracting information about systems in the thermodynamic limit, which often requires finite-size scaling to be performed on simulation results obtained from systems with hundreds (if not many thousands) of spins.  At present, there are no circuit-model quantum computers that can directly simulate spin systems of these sizes.


However, it is well known that system size alone does not determine the classical hardness of simulating a quantum system. While the system size $N$ determines the Hilbert space dimension ($\mathscr{D}\sim e^{N}$), which in turn sets the classical complexity of simulating the system's wave function by brute force, the Hilbert space actually explored by physical systems is highly structured, enabling efficient parameterizations of physical wave functions.  The study of tensor networks over the last few decades has brought this point into sharp focus: Tensor network simulations generally require resources that scale no worse than algebraically with the system size, and only suffer an exponential scaling with respect to the {\it amount of entanglement}, quantified as the bipartite entanglement entropy.  
This realization, and the scaling laws connecting entanglement entropy to equilibrium and non-equilibrium phases of matter, has made it possible to judge by inspection of general properties of a model---e.g. whether it is in equilibrium (and if so if it's at zero or finite temperature), its geometry, its spacial dimension, or its topological properties---whether it can be simulated efficiently on a classical computer or truly requires quantum resources to simulate. 
Situations in which the latter case is realized are compelling examples of hard (and practically relevant) problems for which quantum computing could provide a significant near-term benefit.

In general, the existence of a simple tensor-network representation for a state does not guarantee that properties of that state can be calculated efficiently, because the network may be difficult to contract.  Important examples include when the size (bond-dimension) of the tensors needs to be extremely large to achieve a good approximation, as happens generically for matrix-product state (MPS) simulations of higher-dimensional systems or long-time evolution, or because the tensor network topology does not permit an efficient contraction~\cite{PhysRevLett.98.140506}. In the last few years, several proposals have pointed out that near-term quantum computers may be capable of carrying out tensor-network calculations that are beyond the reach of classical computers \cite{Kim_2017a,Kim_2017b,Kim_2017,Huggins_2019}, with a key insight that the size of that quantum computer can be far smaller than the physical system described by the tensor network.  Very recently, these ideas have been exploited to provide variational energy estimates for the 2D Heisenberg model by simulating small quantum circuits \cite{Liu_2019}.

In this manuscript we present a toolbox for constructing and time-evolving high-bond dimension MPS states on a quantum computer. We refer to these techniques as ``holographic"~\cite{Kim_2017,Kim_2017b} because they enable simulation of a $D$-dimensional system using only a ($D$-1)-dimensional cross-section's worth of qubits by simulating the transfer matrix for the MPS as a quantum channel~\cite{Schon_2005}. The channel effectively moves along the MPS by one unit of distance per channel iteration. Operationally, a purified version of this channel is implemented via unitary operations between a cross-section of spins (physical qubits) and an ancillary quantum memory (bond qubits), followed by partial measurement of the physical qubits. Each iteration of the quantum channel moves one step along the stacking direction of the cross-sections, with physical qubits reset between iterations and reused without duplication, thereby trading spatial resources (qubit number) for time resources (circuit depth).  We present a detailed description of these techniques, and benchmark them via classical simulations of algorithm performance on solvable spin-chains.
 

Next, we show that this representation can actually be made to work by implementing it on a Honeywell trapped-ion quantum computer, and using it to estimate the ground-state energy of the anti-ferromagnetic Heisenberg chain.  A crucial technical ingredient required to perform holographic simulations is the ability to selectively measure and initialize a subset of qubits mid-circuit, without affecting the remaining qubits.  The quantum charge-coupled device (QCCD) architecture~\cite{2020_beta}, in which individual qubits can be dynamically positioned far from other qubits during the execution of a circuit, enables individual addressing (including gates, measurement, and state preparation) with extremely low cross talk, and is therefore very well suited for these types of algorithms. 

Finally, we extend these ideas to the simulation of quench dynamics starting from a holographically generated MPS.  Naively, one would expect the simulation of $N$ initially correlated qubits evolving for a time $t$ under a local Hamiltonian to require a circuit of width $w=N$ and depth $d\sim{\rm poly}(t)$. A constructive algorithm achieving such scaling for $k$-local Hamiltonians was found more than 20 years ago \cite{Lloyd1073},  and in the years since the dependence of the circuit depth $d$ on $t$, $N$, and the error tolerance have all improved (see \cite{Haah_2018,childs2019theory} and references therein).  It might seem that the circuit width requirement $w\sim N$ is fundamental, or even tautological.  While $w=N$ is indeed required in certain worst-case scenarios, in this paper we explore how far fewer than $N$ qubits suffice in many cases of practical interest.  Consider for example a 1D system initially in a state with maximum bipartite entanglement entropy $S$: Time evolution of this state by a local Hamiltonian can be implemented by a circuit of width ${\rm poly}(t)+S$ and depth $N\times{\rm poly}(t)$.  In other words, the required number of qubits is determined by the {\it evolution time} (with a modest constant offset to accommodate the initial entanglement of the state), while the physical size of the system being simulated can be accommodated by increasing the depth of the circuit.  Moreover, if the initial state has a finite correlation length $\xi$, the $N\rightarrow\infty$ limit can be well approximated by a circuit of depth $\sim \xi\times{\rm poly}(t)$.  This reshuffling of resources from circuit width to circuit depth is what we mean by the term ``holographic''.

\section{Holographic simulation of Matrix Product States}
The basis for our quantum simulation algorithms will be the MPS representation of quantum states, which provides an efficient compressed approximation of quantum states with less-than maximal entanglement. We will construct methods for simulating ground-state properties and quench dynamics of local lattice models, defined on a Hilbert space $\mathcal{H}$ that decomposes into a tensor product of ``sites" as $\mathcal{H}=\otimes_{i=1}^\ell\mathcal{H}_i$. Here the sites are arranged along a 1D line with open boundary conditions and have finite local-dimension $|\mathcal{H}_i|=\mathcal{Q}$, and by local we mean that interactions act on at most $k$ sequential sites. Note that this class of systems includes not only 1D spin-chains, where each site simply represents a single spin, but also $D$-dimensional models which can be sliced into a 1D stack of $(D-1)$-dimensional cross-sections (e.g. for a $D$-dimensional cube $\ell = N^{1/D}$).

MPS describe quantum states on such 1D stacks using far-fewer parameters: $\sim \mathcal{O}(\ell\mathcal{Q}\chi^2)$, with $\chi$ the bond dimension---than the worst-case $\mathcal{Q}^{\ell}$ required to specify a generic state, and describe a very special class of states with entanglement entropy across any cut bounded above by $\log \chi$. For ground states of 1D or quasi-1D systems, MPS states can be implemented using classical resources that scale at worst polynomially in the desired accuracy and system size, and provide efficient classical algorithms for simulating low-dimensional ground-states and short-time dynamics. However, classical MPS methods fail for 2D and 3D systems, and for longer-time dynamics where substantial amounts of entanglement have been generated. Nevertheless, many of these systems have far less than the maximal amount of entanglement obtained by random states, and the MPS description provides a dramatic compression. Our goal is to devise an efficient method to prepare and time evolve an MPS-representable state on a quantum computer with an economical use of qubit resources, which gives access to MPS with classically inaccessible bond-dimension, while still leveraging the economical MPS representation for states with less-than maximal entanglement.

To set the stage for these algorithms, we first briefly review a few key properties of MPS that are essential to understanding how they can be represented holographically on a small quantum computer. 

\subsection{Brief review of MPS formalism\label{sec:mps_formalism}}
\begin{figure}[!t]
\begin{center}
\includegraphics[width=1.0\columnwidth]{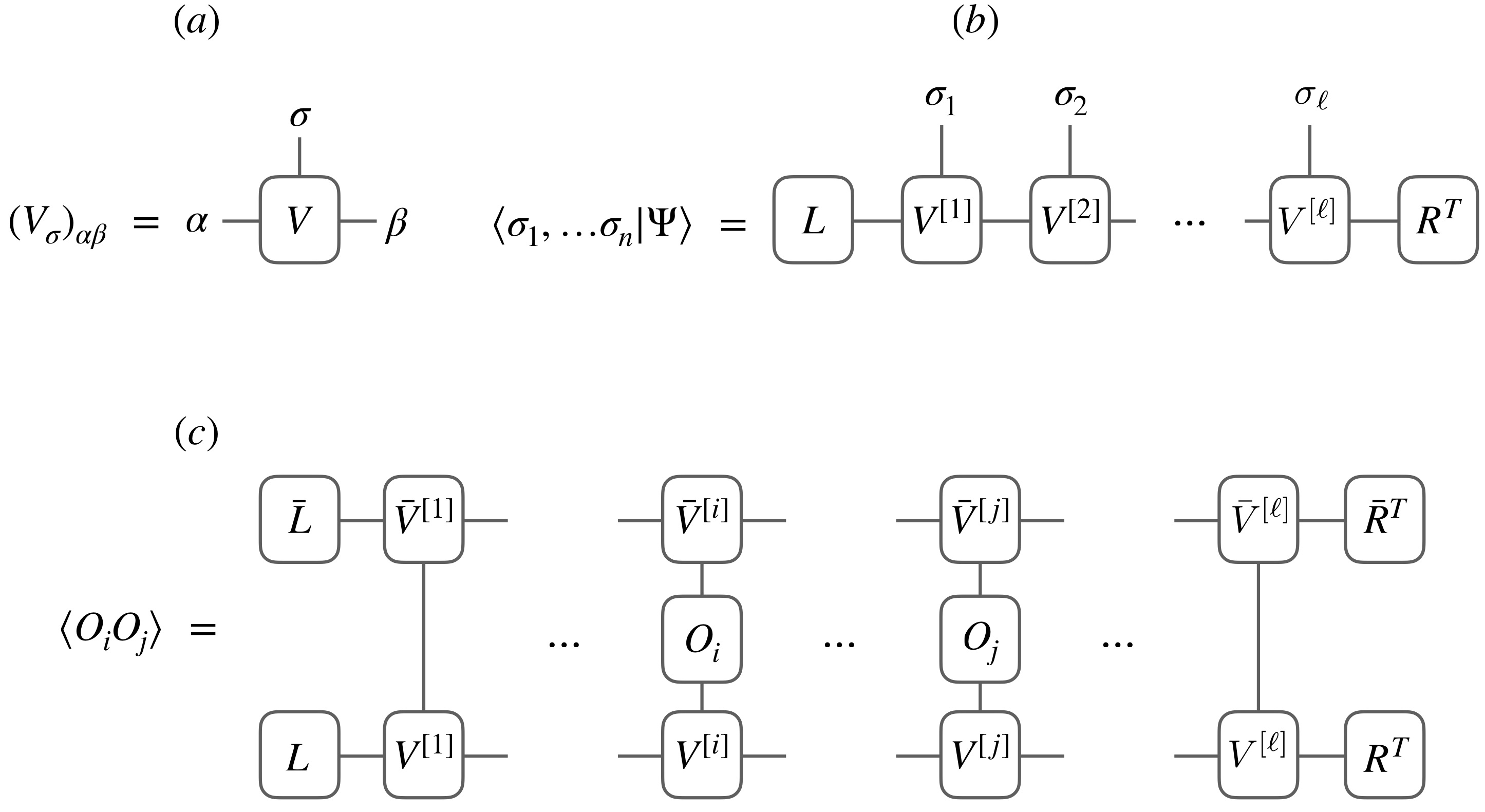}
\caption{Graphical description of an MPS. (a) An individual tensor, (b) the MPS wavefunction as a contraction over such tensors, and (c) the contraction of a tensor network to compute a correlation function.}
\label{fig:MPS_dictionary}
\end{center}
\end{figure}
An MPS with open boundary conditions can be written
\begin{align}
\ket{\Psi}=\sum_{\sigma_1,\dots,\sigma_{\ell}}L^{T}V^{[1]}_{\sigma_{1}}V^{[2]}_{\sigma_{2}}\cdots V_{\sigma_{\ell}}^{[\ell]}R\ket{\sigma_1,\dots,\sigma_{\ell}}.
\label{eq:general_mps}
\end{align}
For each site $j$,  $V^{[j]}_{\sigma_j}$ is a set of $\mathcal{Q}$ square matrices \footnote{In general these matrices can be rectangular, and the bond dimension can depend on $j$.  In this case, we can embed every matrix in a larger square matrix with dimension ${\rm max}_j\chi_j$.}, or equivalently a rank-3 tensor with ``physical'' index $\sigma_j = 1\dots \mathcal{Q}$ and ``bond" indices $\alpha,\beta=1\dots \chi$, where $\chi$ is referred to as the bond dimension. The $\chi$-dimensional vectors $L$ and $R$ specify the left and right boundary conditions, respectively.  The standard graphical representation of an MPS is shown in \fref{fig:MPS_dictionary}. An individual tensor is drawn as a box with a leg for each index, as in \fref{fig:MPS_dictionary}a.  Joined legs imply contraction of the associated tensor indices, such that \fref{fig:MPS_dictionary}b gives the wave function components of $\ket{\Psi}$ [the contractions are implied as matrix-matrix or matrix-vector multiplication in \eref{eq:general_mps}]. 

\begin{figure}[!t]
\begin{center}
\includegraphics[width=1.0\columnwidth]{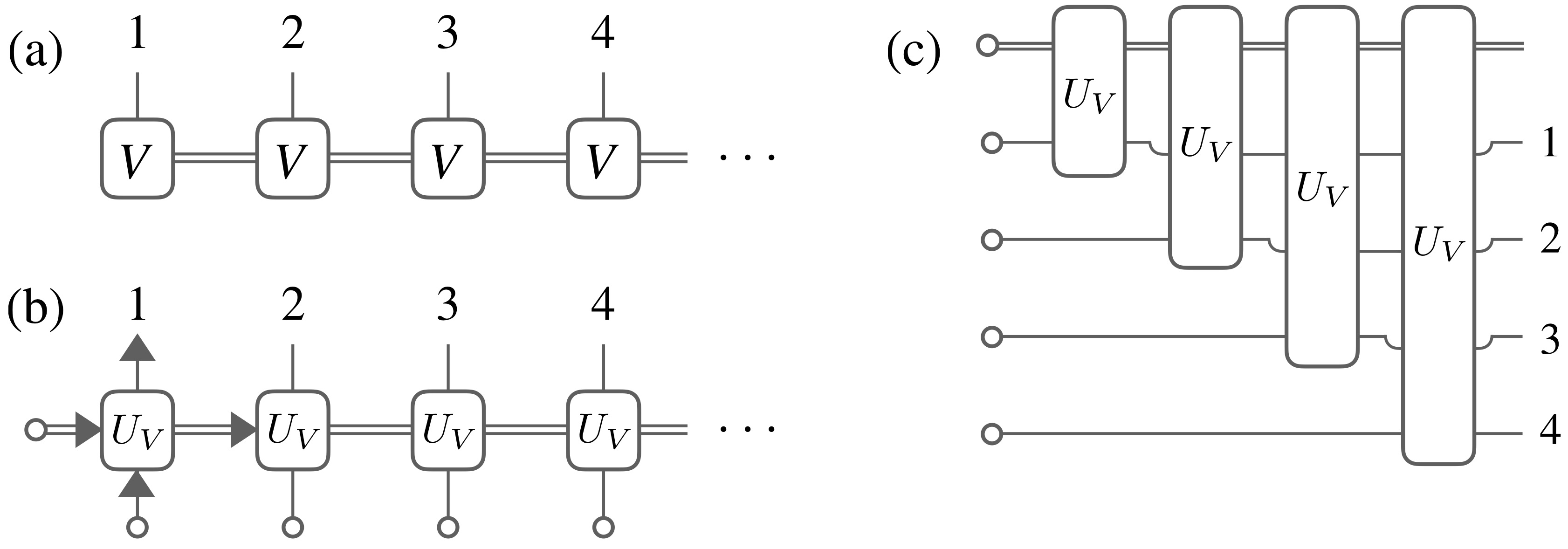}
\caption{MPS as a quantum circuit.  (a) An MPS, assumed to be in right canonical form.  The right canonical condition [\eref{eq:RCF}] guarantees that $V$ is an isometry from $\mathbb{C}^{\chi}\rightarrow\mathbb{C}^{\chi}\otimes \mathbb{C}^{\mathcal{Q}}$, and it can therefore be embedded in a unitary $U_V$ acting on $\mathbb{C}^{\chi}\otimes \mathbb{C}^{\mathcal{Q}}$ but restricted to a fixed input of the physical qubit (denoted with an open circle), as in (b). The unitary evolution in (b) is equivalent (up to simply rearranging the lines) to the circuit diagram in (c).}
\label{fig:MPS_schematic}
\end{center}
\end{figure}

One can imagine creating an $\ell$-site MPS as a physical state of a quantum computer by letting an ancilla register (containing ``bond qubits'') interact unitarily and sequentially with $\ell$ physical registers \cite{Schon_2005}, each representing one site of the MPS, as in \fref{fig:MPS_schematic}(a-c).  In these circuit diagrams and elsewhere, open circles denote initialization of a qubit (or register of qubits) to the $\ket{0}$ state. To implement this construction with unitary circuit elements, one must exploit the gauge redundancy of the MPS description \cite{perez2006matrix} to place the MPS in {\it right canonical form} (RCF) such that
\begin{align}
\sum_{\sigma}V^{[j]\phantom\dagger}_\sigma\! V^{[j]\dagger}_{\sigma}= \mathds{1}~~~~~ \forall j.
\label{eq:RCF}
\end{align}
The relationship between RCF and unitary embedding will be discussed further in the next section, but for now we simply want to emphasize that the aforementioned MPS gauge redundancy ensures that we can, without loss of generality, assume that \eref{eq:RCF} holds for the state in \eref{eq:general_mps}. Note that this definition of RCF is slightly non-standard: the boundary tensors are typically ordinarily also canonical, however we choose this version of RCF because it simplifies much of what follows.  While this choice imposes some limitations on how faithfully the right-boundary condition can be imposed in a quantum circuit, bulk physics will be unaffected
Note also that for a normalized state $\ket{\Psi}$, the imposition of RCF implies that in general both $L$ and $R$ cannot be simultaneously normalized.  Without loss of generality we take $L$ to be normalized but not necessarily $R$.
 
 \begin{figure*}[!t]
\begin{center}
\includegraphics[width=1.5\columnwidth]{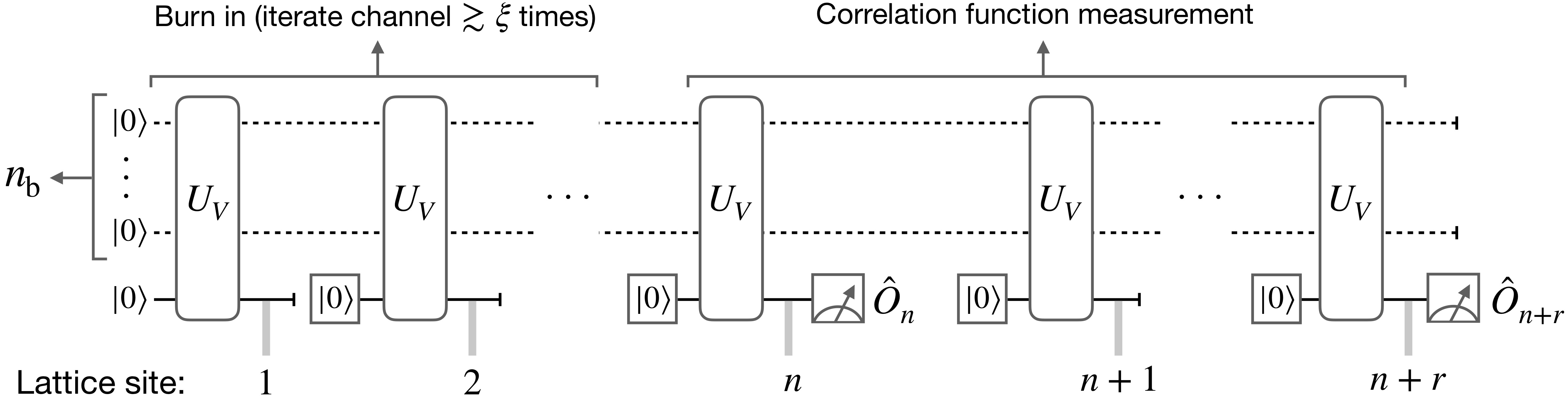}
\caption{A circuit implementing holographic state preparation and correlation function measurement of an MPS.  In this simple example we use one system qubit (enabling simulation of an infinite 1D spin-$\nicefrac{1\!\!\;}{\!\!\;2}$ chain) and $n_{\rm b}$ bond qubits for an MPS with bond-dimension $\chi=2^{n_{\rm b}}$.  Open circles denote qubit initialization to the $\ket{0}$ state, and the measurements are implied to be in the eigenbasis of the operators $\hat{O}_n$ and $\hat{O}_{n+r}$. Averaging over the outcome of this circuit (with the two measurement results multiplied together) produces the correlation function $\langle\hat{O}_n\hat{O}_{n+r}\rangle$.}
\label{fig:holo_state_prep}
\end{center}
\end{figure*}
 
Correlation functions of local operators, $\<\hat{O}_i\hat{O}_j\>$, can be computed by contracting the tensor network depicted in \fref{fig:MPS_dictionary}c. An efficient method to contract such a network for a long chain is to first contract the physical legs of the tensors on each site to form transfer matrices
\begin{align}
\mathcal{E}^{[k]}_{\alpha\gamma,\beta\delta} = \sum_{\sigma_k}  \big(\bar{V}^{[k]}_{\sigma_k}\big)_{\gamma,\delta} \big(V^{[k]}_{\sigma_k}\big)_{\alpha\beta}
\end{align}
on each site $k\neq i,j$ (with $\bar{V}$ the complex conjugate of $V$), and begin multiplying these transfer matrices from left to right.  When site $i$ is encountered, we apply the modified transfer matrix
\begin{align}
\mathcal{O}^{[i]}_{\alpha\gamma,\beta\delta}= \sum_{\sigma_i,\tau_i}\big(\bar{V}^{[i]}_{\tau_i}\big)_{\gamma\delta} \bra{\tau_i}\hat{O}_i\ket{\sigma_i} \big(V^{[i]}_{\sigma_i}\big)_{\alpha\beta},
\end{align}
and similarly for site $j$. If we interpret the bond vector space as the Hilbert space of some quantum system, then the transfer matrix is a linear superoperator acting on bond-space density matrices $\rho_{\beta\delta}$ $(\beta,\delta\in\{1,2,\dots \chi\})$ via the mapping
\begin{align}
\mathcal{E}\!: \rho\rightarrow \sum_\sigma V^{\dagger}_\sigma\rho V_\sigma^{\phantom\dagger}.
\label{eq:tchannel}
\end{align}
Together with the RCF conditions in \eref{eq:RCF}, \eref{eq:tchannel} establishes $\mathcal{E}$ as a quantum channel (trace-preserving completely positive map) on the bond space \cite{perez2006matrix}, with the MPS matrices $V_\sigma$ as the Krauss operators of the channel.  In this language, the contraction depicted in \fref{fig:MPS_dictionary}c can be expressed as the overlap of a ``time-evolved'' initial bond-space density matrix $\rho_{\rm i}=|L \rrangle \llangle L|$
(with ket $|\,\dots\rrangle$  indicating a state in the bond Hilbert space) with the final un-normalized state $|R\rrangle$,
\begin{align}
\label{eq:channel}
\langle \hat{O}_i\hat{O}_j\rangle\!=\!\llangle  R|\mathcal{E}^{[\ell]}\!\circ\cdots\mathcal{O}^{[j]}\!\circ\cdots\mathcal{O}^{[i]}\!\circ\cdots\mathcal{E}^{[1]}(\rho_{\rm i})|R\rrangle.
\end{align}
Indeed, the earliest comprehensive treatment of MPS in the literature (in which they were called {\it finitely-correlated states}) defined them in terms of quantum channels \cite{Fannes_1992}.

\subsection{Holographic MPS generation}
Equation \ref{eq:channel} demonstrates that correlation functions of an MPS can be encoded in the dynamics of a quantum system with size independent of $\ell$; the {\it spatial} structure of the physical Hilbert space has been converted into a (discrete) {\it time} direction of the bond Hilbert space.  Thus one can simulate the state of a $D$-dimensional system using a system of dimension $D-1$, inspiring the moniker ``holographic"~\cite{Kim_2017a}.  However, it is important to keep in mind that this dynamics is not unitary. The holographic algorithms described here can be viewed as explicit purifications of this non-unitary dynamics in the form of quantum circuits.  Alternatively, one can understand these algorithms starting from the known representation of MPS as quantum circuits, previously described in Sec.\,\ref{sec:mps_formalism} and illustrated in \fref{fig:MPS_schematic}.  From this perspective the dimensional reduction can be understood by looking at the causal structure of that circuit and recognizing that the physical register corresponding to site $j$ can be measured and reset before the bond qubit register interacts with the physical register of site $j+1$, implying that only a single site worth of physical qubits is required to implement the entire circuit \cite{Huggins_2019}. Despite the constant erasure of information in the physical qubits, long-range spatial correlations in the system are retained as memory in the bond qubits.

Generic versions of such holographic algorithms for 2D systems were previously outlined in Ref.\,\cite{Kim_2017a}, though without explicit discussion of connections to the MPS formalism, and the known representation of MPS as quantum channels. Subsequent work \cite{Kim_2017b,Kim_2017} revealed an intriguing element of noise resilience in holographic simulation techniques. Due to the repeated partial-measurement and reset in the holographic technique, errors do not propagate indefinitely as for purely unitary circuits. Consequently, a finite density of errors produces a finite imprecision on the measured correlation function, in contrast to a purely unitary circuit, for which a single error can spread and contaminate all outputs.

In what follows, we unify these perspectives with the framework of MPS, and develop concrete variational ground-state preparation and quantum dynamics simulation techniques using this framework. We begin with a detailed description of the holographic MPS preparation/measurement protocol, summarized in \fref{fig:holo_state_prep}. This protocol utilizes a register of $n_{\rm b}=\log_2\chi$ ``bond" qubits (representing the $\chi$-dimensional bond Hilbert space) initialized in state $|L\rrangle$, and  a register of $n_{\rm p}=\log_2 \mathcal{Q}$ ``physical"-qubits (representing the $\mathcal{Q}$-dimensional physical Hilbert space of a single lattice site) prepared in a fixed reference state $|0\>$.   The channel $\mathcal{E}$ is realized by applying unitary gates between the physical qubits and bond qubits, and then tracing out (i.e.\ discarding) the physical qubit.  Such a unitary purification of the MPS-channel can always be constructed via the Stinespring dilation.  Specifically, one can embed each MPS tensor $(V_\sigma)_{\alpha,\beta}$ as the columns of a unitary matrix $U_{\alpha,\sigma; \beta,\sigma'}$ with fixed index $\sigma'=0$, i.e. $(V_\sigma)_{\alpha,\beta} = \<\sigma|\llangle\alpha|U|0\>|\beta\rrangle$. Since, in right canonical form, $V$ forms an isometry from $\mathbb{C}^\chi \rightarrow \mathbb{C}^{\chi}\otimes \mathbb{C}^{\mathcal{Q}}$, the columns with $\sigma'=0$ form an orthonormal set, which can always be completed into a full orthonormal basis for $\mathbb{C}^{\chi}\otimes \mathbb{C}^{\mathcal{Q}}$ to obtain $U$.

Any correlation function, $C=\<\Psi| \big(\otimes_i\hat{O}_i \big)|\Psi\>$, of the corresponding MPS can be sampled by iterating this quantum channel to step through the sites of the chain from $1$ to $\ell$ in the following sequence of steps:
\begin{enumerate}
\item {\bf State prep:} Start with $C=1$. Prepare the bond-qubit register in a given state $|L\rrangle$, which sets the left-boundary condition for the MPS. Then, starting with site $i=1$:
\item {\bf While $i\leq \ell$:} Iteratively apply the quantum channel for the MPS matrix to step along the chain from site $1$ toward site $\ell$ by the following steps:
\begin{enumerate}
\item Prepare the physical qubit register in a reference state, $|0\>$.
\item Act on the physical qubits and bond qubits with a unitary circuit $U$, which is a purification of the MPS on-site tensor for site $i$.
\item Measure the physical qubits in the eigenbasis of operator $\hat{O}_i$ (which may be the identity operator on most sites, for which the measurement is unnecessary). Denote the eigenvalue of $\hat{O}_i$ corresponding to the measurement outcome by $\lambda_i$, and multiply $C\leftarrow \lambda_i C_i$.
\item Increment: $i\leftarrow i+1$. 
\end{enumerate}
\item Measure the bond-qubits in a basis containing $|R\rrangle$.
\end{enumerate}

If we post-select on outcomes for which the bond qubits are found to be in state $|R\rrangle$, then the above algorithm samples the correlator $C$ in the state $|\Psi\>$, and the expectation value can be estimated by averaging over sufficiently many repetitions to achieve a desired statistical precision~\footnote{For local correlation functions where $\hat{O}_i=\mathds{1}_i$ on all but a fixed number of sites, the variance of the observable should not scale with system size, and statistical accuracy can be achieved with a system-size independent number of repetitions. Non-local observables with weight on extensively many sites may have high-variance and cannot in general be efficiently estimated.}. Note that fixing the right-boundary condition through post selection incurs a multiplicative overhead $\sim\chi$, which will be very large in cases where quantum computation is required. However, since we are primarily interested in bulk properties, and since correlations decay exponentially in distance from the boundary, it is generally sufficient to skip this post-selection, which provides a weighted average of the correlation function over the right-boundary condition.  If we are interested in boundary effects, for example when examining impurity models or boundary conformal field theories, one can study the left boundary (where the boundary condition is set by the state-preparation for the bond-register).

\subsection{Holographic entanglement measurements}
Measures of entanglement provide detailed insights into quantum many-body systems beyond what can be drawn from local correlation functions, revealing non-local correlations, topological and symmetry-protected topological orders~\cite{levin2006detecting,PhysRevLett.96.110404,pollmann2010entanglement} , and diagnosing thermalization, scrambling, and many-body localization~\cite{abanin2019colloquium}.  Methods to measure R\'enyi entropies
\begin{align}
S_A^{(n)} = \frac{1}{1-n}\log \text{tr}\rho_A^n 
\end{align}
for a subsystem $A$, with integer index $n$, have been developed based on creating replica copies of the system and measuring operators that cyclically permute the quantum states of various copies \cite{ekert2002direct}, or by examining cross-correlations in randomized measurements to virtually implement the desired replicas \cite{brydges2019probing}. These methods can be directly adapted to holographically represented states by performing the desired measurements on the physical qubit registers, as described in the previous section. Such measurements would enable, for example, the estimation of free energy from holographically generated thermofield double states \cite{wu2019variational,zhu2019variational}.

If one is interested in the entanglement entropy of a bipartition of the chain, it can be directly obtained from measurements of the bond-qubit register. Namely, since the holographic simulation method recreates an MPS in right canonical form, the entanglement spectrum for the physical qubits bipartitioned by cutting between sites $j$ and $j-1$ is precisely equal to the spectrum of the density matrix for the bond-qubits after $j$ iterations of the holographic simulation algorithm. In this case, the replica-SWAP or randomized-measurement techniques can be applied directly to the bond qubits, with a number of measurements that grows with $\chi$, but not the interval size, offering a potential savings in measurement complexity. For replica-SWAP based measurements of entanglement entropy, this holographic method provides a potentially huge savings in qubits required, as one needs only to replicate the bond-qubits, without replicating an extensive number of physical qubits for every site in the chain.

\subsection{Expressivity of holographic MPS}
While a unitary circuit representing the (purified) quantum channel of any MPS is formally guaranteed to exist, the crux for practical use of holographic simulation techniques will be constructing effective methods for implementing channels for physically relevant systems using low-depth circuits. Namely, arbitrary unitary synthesis from a local gate set generically requires gate counts that scale exponentially with qubit number, and is clearly not a viable technique for large bond dimension. However, physical systems with local Hamiltonians are far from ``generic", and have considerable structure that could be exploited for efficient simulation. This observation poses the following basic question: What class of quantum states can be efficiently holographically represented on a quantum computer with exponentially large $\chi\sim 2^{n_{\rm b}}$, but using low-depth [i.e.\ with $\text{poly}(n_{\rm b})$ gates] quantum circuits? Though we cannot conclusively answer this general question, in the following we develop holographic algorithms for simulating non-equilibrium dynamics starting from correlated ground states, and provide numerical evidence demonstrating that low-depth circuits may suffice in many situations of practical interest.

\section{Holographic Variational Quantum Eigensolver  (\lowercase{holo}VQE)\label{sec:holo_vqe}}

A central task for quantum materials simulation is to accurately approximate the ground-state correlations of local Hamiltonians $H=\sum_j h_j$, where each term $h_j$ acts on sites within a distance at most $k$ from site $j$.  Hybrid classical/quantum variational algorithms, like the variational quantum eigensolver (VQE) \cite{Peruzzo:2014aa}, offer promising methodologies for attacking this problem on moderate scale quantum computers.  In VQE, one prepares a trial wave-function $|\psi(\theta)\>$ on a quantum computer by evolving a fixed initial state with a quantum circuit composed of gates parameterized by rotation angles $\theta \in \mathbb{R}^p$  ($p$ being the number of variational parameters). The expectation value of the energy, $E(\theta)=\sum_j \<\psi(\theta)|h_j|\psi(\theta)\>$, is subsequently estimated by measuring each individual term in the sum to the desired precision. Then, a classical computer updates the parameters $\theta$ to lower the variational energy in order to find the best approximation of the true ground state within the family of states, $|\psi(\theta)\>$.

The holographic representation of MPS on a quantum computer naturally suggests a holographic extension of VQE (holoVQE), which uses the MPS representation described in the previous section with the unitary $U_{V}$ represented by a parameterized circuit. The expectation value of energy can be computed by measuring each term in the Hamiltonian using the above-described procedure for measuring correlation functions. Then, the variational ansatz can be optimized by using a classical algorithm to minimize $E(\theta)$.

The implementation of holoVQE is simplified for crystalline materials with translation invariant Hamiltonians, where there are only a finite number of terms in the Hamiltonian that must be independently measured in the thermodynamic limit. The holographic method produces an MPS with open boundary conditions which is not translation invariant. However, in an MPS the boundary's influence decays exponentially with the correlation length $\xi$, so one can simply measure the distinct terms in the Hamiltonian a distance $r\gtrsim \xi$ from the boundary. In the holographic correspondence, we move a large distance into the spatial bulk by iterating the MPS quantum channel for a long time to burn in its steady state, as in \fref{fig:holo_state_prep}.

In the following sections, we demonstrate a simple application of the holoVQE technique for approximating the ground-state energy of an XXZ chain using only a single ancillary bond qubit, and then physically implement this technique on a trapped ion quantum computer to analyze the SU$(2)$-symmetric Heisenberg point. We show that symmetry principles can be incorporated into the variational circuit ansatz to reduce the number of variational parameters and simplify the optimization. Next, we consider a model with lower symmetry: the transverse-field Ising model (TFIM). We apply a more generic circuit ansatz, and analyze the scaling of algorithm performance for ground-state preparation and reconstruction of critical correlation functions with increasing number of bond qubits. For small circuit sizes, we are able to reproduce the optimal MPS ansatz at bond dimension $\chi=2^{n_{\rm b}}$, obtaining relative accuracy on the (infinite size) critical ground-state energy below $10^{-4}$ with only a few qubits.


\subsection{Role of symmetries}
The crux of effectively implementing a variational procedure is constructing a good variational ansatz. For present purposes, this entails identifying a parameterized unitary circuit family that effectively implements the purified transfer matrix of the desired MPS approximation to the ground state. Consideration of symmetries can guide the design of these circuits.

Symmetries of the model can be strictly enforced on the variational states by restricting attention to symmetry preserving circuit families. Here, by symmetry-preserving circuit, we mean that we choose a particular linear representation of the symmetry action on the bond qubits (possibly including projective representations~\cite{pollmann2017symmetry} when dealing with potential symmetry-protected topological states), and ensure that parameterizations of the variational circuit preserve the total symmetry quantum numbers of the physical and bond qubits together.

We note that it is not always desirable to explicitly enforce all symmetries of the model. For example, we may wish to assess whether or not the ground state spontaneously breaks symmetries, in which case one could build an ansatz around various possible symmetry-broken configurations. Moreover, it is often the case that, for a fixed bond dimension, the lowest energy state may not preserve the full set of symmetries possessed by the ground state.  For example, when $\chi=1$ (no bond qubits) an MPS simply corresponds to a mean-field (best product-state) ansatz, for which energy minimization often yields symmetry-broken solutions. The examples below are indicative of these various possibilities.

\subsection{XXZ chain}
The XXZ spin chain is a canonical model of strongly correlated 1D systems, describing both one-dimensional quantum magnetism and superfluidity (by mapping the spins to hard-core bosons).  For nearest-neighbor interactions the Hamiltonian is
\begin{align}
H = J\sum_i \(X_iX_{i+1}+Y_iY_{i+1}+\Delta Z_iZ_{i+1}\),
\end{align}
where (without loss of generality) we take $J>0$.  The model has a global U$(1)$ symmetry, enlarged to a full SU$(2)$ symmetry at the Heisenberg points $\Delta=\pm 1$.  For $\Delta<-1$ [$\Delta>1$] the spectrum is gapped and the symmetry-broken ground state is ferromagnetically [antiferromagnetically] ordered, while for $|\Delta|\leq 1$ the model is gapless and has no long-range spin order.  For all values of $\Delta$ the model is exactly solvable by Bethe ansatz \cite{faddeev1995algebraic}.
\begin{figure}[t]
\begin{center}
\includegraphics[width=0.95\columnwidth]{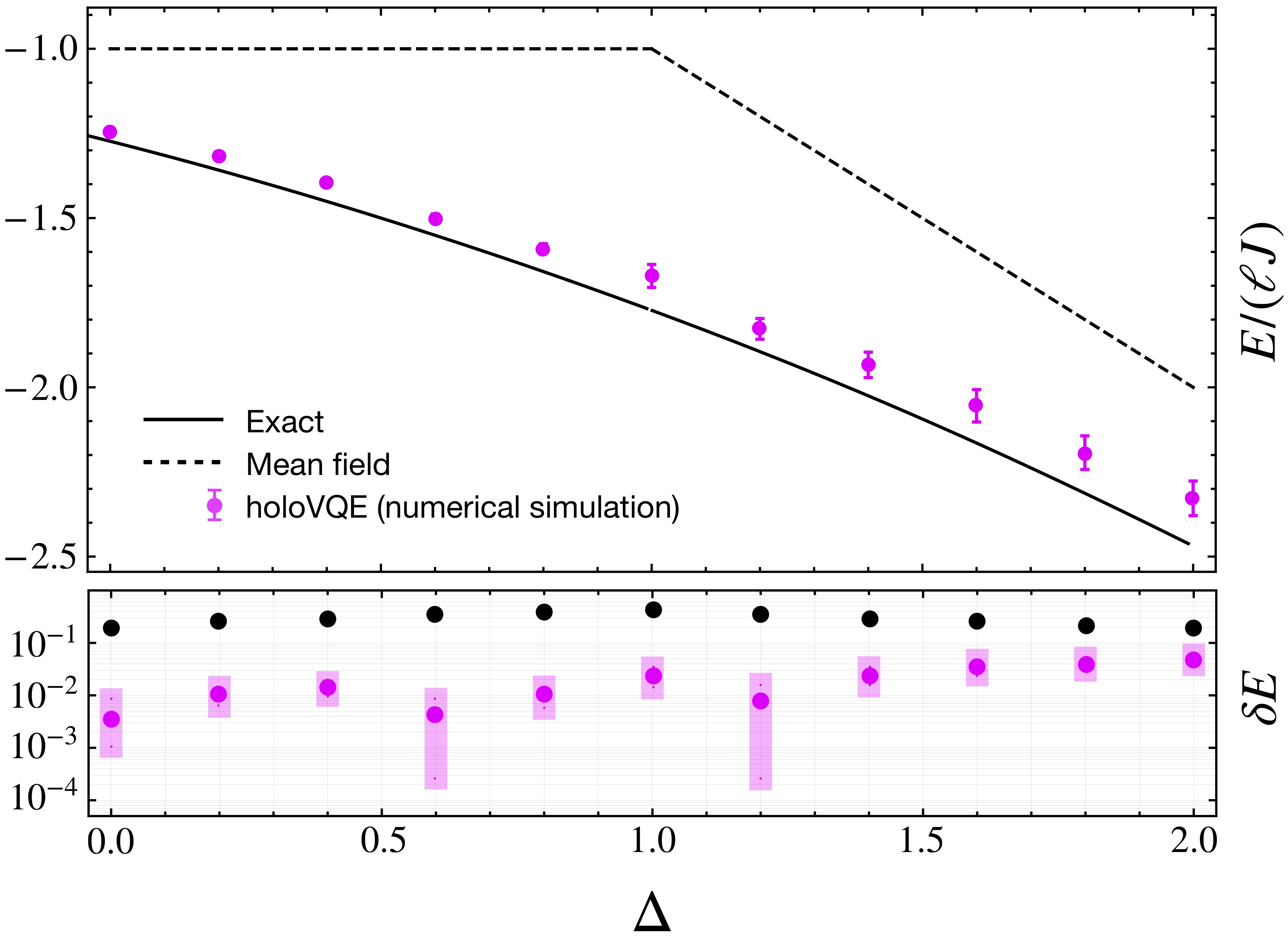}
\caption{Top: Energy per site of the XXZ chain in the thermodynamic limit.  The dashed line is mean-field theory, the solid line is the exact energy obtained from Bethe ansatz, and the points are from simulated holoVQE using a single bond qubit (error bars represent $1\sigma$ sampling uncertainties).  Bottom: Fractional energy error [$\delta E=|(E-E_{\rm exact})/E_{\rm exact}|$]  of holoVQE (pink points with error bars) and mean-field theory (black points), showing that the addition of even a single bond qubit can drastically improve the ground-state energy estimate.
}
\label{fig:xxz}
\end{center}
\end{figure}
%
%
%
 

In the antiferromagnetic phase ($\Delta\geq 1$) the mean-field solution is antiferromagnetically ordered, spontaneously breaking discrete-translational symmetry (and SU(2) symmetry for $\Delta=1$).  Since this state is consistent with the known value of $S_z=0$ for the true ground state, we can build a $\chi=2$ MPS by introducing a single bond qubit, and allowing it to interact with the system qubit via unitaries that conserve total $S_z$ (note that by breaking discrete translational symmetry, this $\chi =2$ MPS achieves the same energy as a $\chi=4$ translationally invariant MPS).  Choosing $U=\exp[-i\theta(X_{\rm p}X_{\rm b}+Y_{\rm p}Y_{\rm b})]\exp[-i\phi(Z_{\rm p}Z_{\rm b})]$, we can now use a holographic representation of the MPS to measure the energy for a given choice of parameters $(\theta,\phi)$, and minimize using a classical feedback loop.  Using gradient descent and simulating 256 shots per energy measurement, we obtain the results shown for $\Delta\geq 1$ in \fref{fig:xxz}.

Additional care must be taken when computing the energy for $0<\Delta<1$.  In this case, the mean-field ground state breaks the U(1) symmetry of the model by spontaneously aligning (antiferromagnetically) along some direction in the XY plane.  Since this state does not live in the correct symmetry sector with respect to U$(1)$, it is not sufficient to restrict our attention to circuits that conserve total $S_z$.  In these cases, we find that the idea MPS at $\chi=2$ can be obtained using the three-parameter ansatz $U=\exp[-i(\theta X_{\rm p}X_{\rm b}+\phi Y_{\rm p}Y_{\rm b}+\psi Z_{\rm p}Z_{\rm b})]$.


\begin{figure}[t]
\begin{center}
\includegraphics[width=1.0\columnwidth]{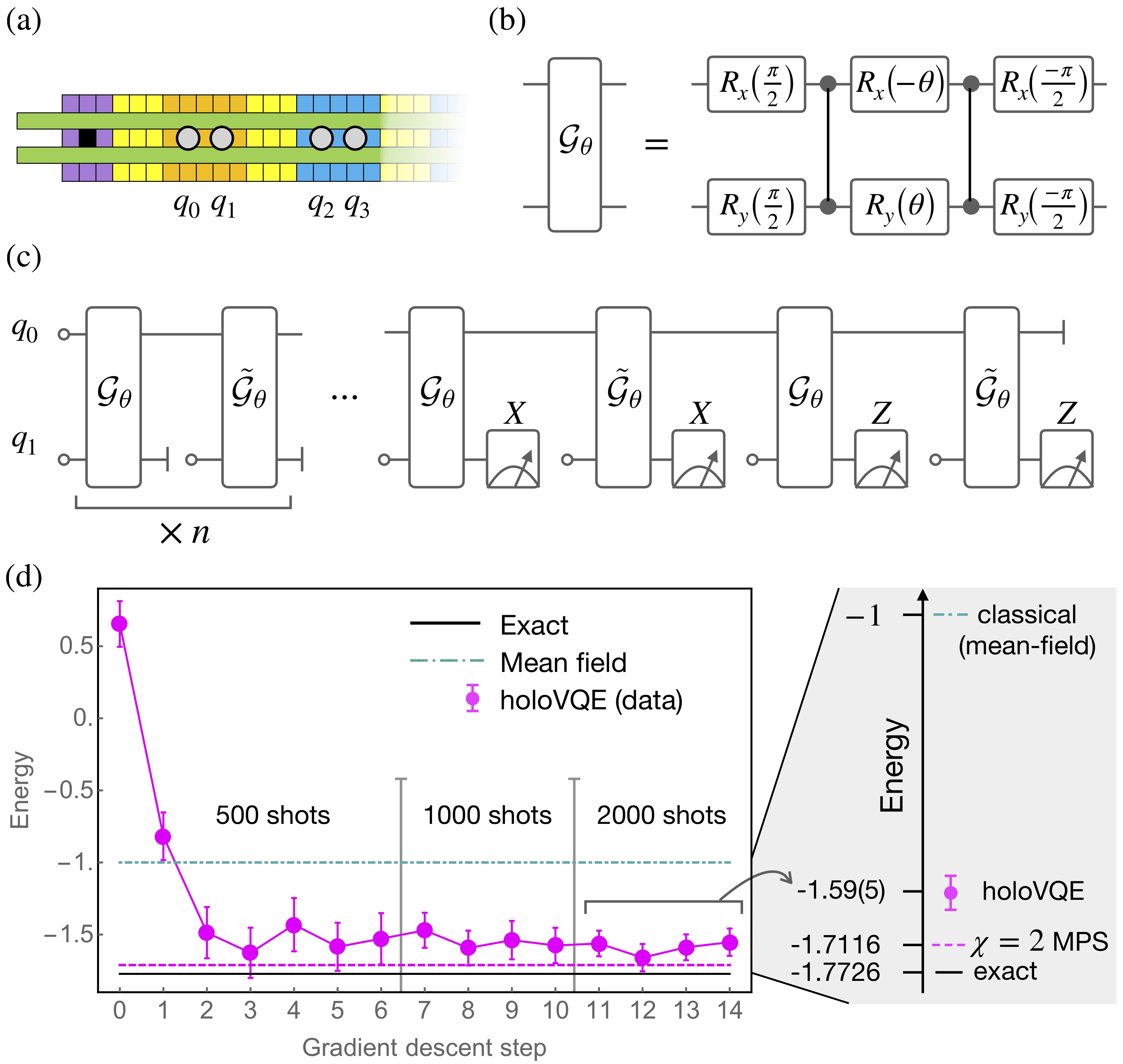}
\caption{Experimental implementation of holoVQE on a Honeywell trapped-ion quantum computer.  (a) Schematic of the ion trap; for this experiment we used two neighboring gate zones, each loaded with 2 data qubits and two sympathetic cooling ions (not shown).  (b) Decomposition of $\mathcal{G}_{\theta}$ into two ${\rm c}Z$ gates. (c) The circuits used for holoVQE involve one physical qubit and one bond qubit, though we utilize the two gate zones to parallelize the data taking. (d) Representative data from holoVQE.}
\label{fig:expt_results}
\end{center}
\end{figure}

\subsection{Trapped ion implementation: Heisenberg chain}
At the antiferromagnetic Heisenberg point, we implement the holoVQE procedure experimentally using Honeywell's QCCD trapped-ion quantum computer described in Ref.\,\cite{2020_beta}.  We utilize a subset of the 5 designated ``gate zones'' (orange/blue in \fref{fig:expt_results}a), which suffices to run two parallel instances of the holographic state preparation protocol with a single bond qubit for each \footnote{running two instances in parallel simply allows us to achieve smaller error bars for a given total run time.}.  At the Heisenberg point, it can be shown that the two-parameter ansatz described above for $\Delta\geq1$ is actually unnecessarily flexible, and it suffices to restrict our attention to $\phi=0$.  Thus the entangling unitary between physical- and bond-qubit is 
\begin{align} 
	\mathcal{G}_{\theta} = \exp[-i\theta(X_{\rm p}X_{\rm b}+Y_{\rm p}Y_{\rm b})/2].
\end{align}
The native two-qubit gate for our architecture (the M\o lmer S\o rensen gate \cite{PhysRevLett.82.1835}) is local-unitary-equivalent to a controlled-$Z$ (c$Z$) gate, at least two of which are necessary to synthesize $\mathcal{G}_{\theta}$ for arbitrary $\theta$ (a minimal decomposition is shown in \fref{fig:expt_results}b).  The holographic MPS circuit is then built by alternating applications of $\mathcal{G}_{\theta}$ and $\tilde{\mathcal{G}}_{\theta}=X_{\rm p}\mathcal{G}_{\theta}$ (this alternation corresponds to starting with a classical antiferromagnet, as discussed above), with reinitialization of the physical qubit in between each entangler (see \fref{fig:expt_results}c). For the small bond dimension accessed in this example, we find that a ``burn in'' distance of 4 lattice sites is sufficient to approximate bulk expectation values to well within shot noise. Figure \ref{fig:expt_results}d shows the results of holoVQE.  Starting with a randomly chosen parameter $\theta$, we use gradient descent with derivatives estimated from finite differences of the measured energies $E(\theta)$, increasing the shot count for each energy measurement from $500$ to $2000$ as gradient descent proceeds.  Error bars are $2\sigma$ confidence intervals obtained from a non-parametric bootstrap resampling of the data.  Averaging over the final four data points (taken at the highest shot counts) we obtain an estimate of $E=-1.59(5) J$ for the per-site ground-state energy. For comparison, the mean-field ground state---which lowers the energy as much as possible without entanglement---achieves $E=-J$ (blue dot-dashed line in \fref{fig:expt_results}d).  The optimal MPS with bond dimension $\chi=2^{n_{\rm b}}=2$ achieves $E\approx-1.712 J$, which the experimental results would converge to in the absence of noise or other imperfections (purple dashed line in \fref{fig:expt_results}d), while the exact ground-state ($\chi = \infty$) has $E=J(1-4\log 2)\approx -1.773 J$ (from Bethe ansatz, black solid line in \fref{fig:expt_results}d).

Note that this measured value provides a proper variational upper bound for the {\it infinite chain}, despite being obtained from a small quantum circuit.  To highlight the resource savings of holoVQE, we note that achieving comparable accuracies using brute-force simulation of an $L$-site chain would require $L=6$ (rather than 2) qubits to sufficiently suppress finite-size effects. If the circuit infidelities were reduced by either improving the gate fidelities or using error mitigation techniques, the minimum achievable energy with 2 qubits for holoVQE, $E_{\rm min}\approx -1.712J$, would require 10 (perfect) qubits to achieve by brute force.

\subsection{Increasing the bond dimension: Transverse field Ising model (TFIM)}
In the previous examples, we used just a single bond qubit corresponding to MPS with bond-dimension two. These results already demonstrate the dramatic compression of resources enabled by the holographic simulation method in achieving reasonable accuracy on an infinite, critical spin-chain using only a pair of qubits. However, turning holoVQE into a useful algorithm requires a method to systematically improve the accuracy of the simulations. We now explore the performance of holoVQE upon including additional bond qubits, focusing on the task of ground-state energy estimation for the 1D transverse-field Ising model (TFIM), with Hamiltonian
%
\begin{align}
H_\text{TFIM}=-\sum_j \( J Z_{j}Z_{j+1}+h X_j\).
\end{align}
The TFIM exhibits a ground-state phase transition from an ordered ($h<J$) to a disordered ($h>J$) phase, both of which are gapped and can be well described by MPS of fixed (system-size independent) bond dimension. These phases are separated by a self-dual critical point at $J=h$, described by a conformal field theory (CFT) with central charge $c=1/2$, whose non-constant entanglement scaling requires a bond dimension that grows with system size as $\chi\gtrsim L^{c/3} = L^{1/6}$ to achieve asymptotically accurate correlations. Nevertheless, it turns out that the ground-state energy and moderate-range spin-correlation functions of this model can be captured with fairly high accuracy using modest bond dimension MPS, even at the critical point \footnote{We note that the anomalous exponent $\eta$ controlling the decay of the longitudinal spin correlations is an exception to the above rule, and is only qualitatively captured for the small bond dimensions explored here.}.
\begin{figure}[t]
\begin{center}
\includegraphics[width=0.95\columnwidth]{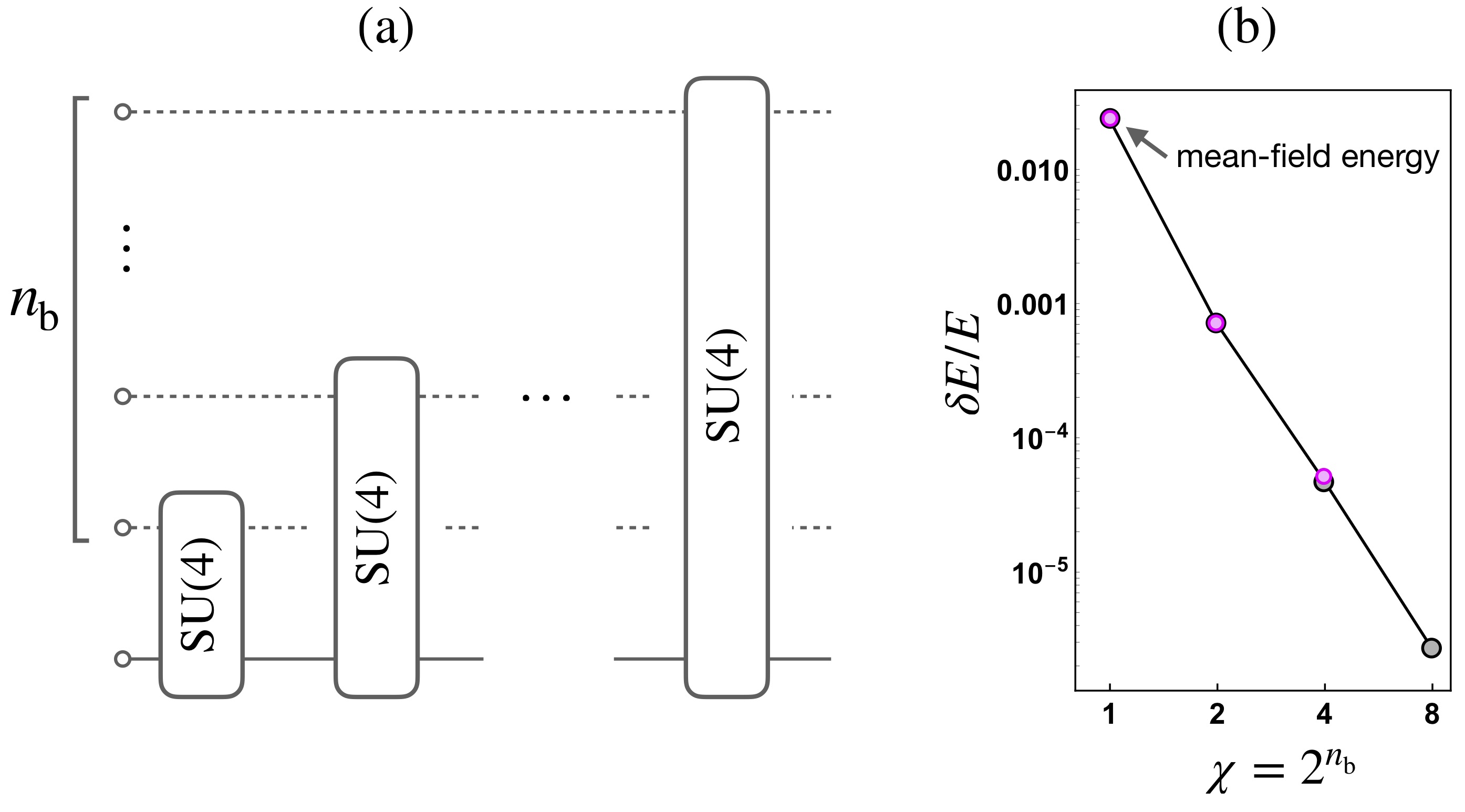}
\caption{(a) Circuits used for holographic MPS representation of the TFIM ground state.  Each arbitrary SU(4) is decomposed into three native two-qubit gates and 8 single-qubit gates, with a total of 15 real variational parameters (as in Ref. \cite{hanneke_2010}).   (b) Energy obtained using ``star'' circuits with an increasing number of bond qubits. The agreement with exact MPS energy minimization at bond dimension $\chi=2^{n_{\rm b}}$ is excellent, although we were not able to obtain reliable energies from this ansatz for $n_{\rm b}\geq 3$.}
\label{fig:TFIM}
\end{center}
\end{figure}

To explore the efficacy of holoVQE for this paradigmatic toy model, we numerically simulate the holoVQE procedure at the critical point ($h=J$) for a sequence of variationally parameterized circuits with a variable number of bond qubits. For $n_{\rm b}$ bond-qubits, we construct a ``star" circuit that involves only two-qubit gates that sequentially entangle the physical qubit and the $j^\text{th}$ bond-qubit for $j=1,\dots n_{\rm b}$, allowing each individual gate to be an arbitrary $\in {\rm SU}(4)$ two-qubit unitary, as in \fref{fig:TFIM}a.  The primary challenge in these calculations is to reliably find the global minimum of a constrained non-linear optimization problem; we were only able to find reliable results for $n_{\rm b}=0,1,2$, for which simulated annealing worked well.  We note that classical MPS calculations also suffer from this challenge, which is typically overcome by breaking translational invariance in order to make the problem linear (at the cost of greatly expanding the parameter space).  It is clear that scaling holoVQE to large circuits (and therefore large effective bond dimension) will require significant further development along these lines.

The results obtained by brute-force global optimization are shown in \fref{fig:TFIM}b, along with those obtained by an unconstrained MPS optimization using bond-dimension $\chi=2^{n_{\rm b}}$.
Surprisingly, this simple circuit design finds the best possible MPS even for $n_{\rm b}=2$, for which the parametrization is not exhaustive of all $n_{\rm b}+1=3$ qubit unitaries.  Because global optimization strategies did not yield reliable improvements for $n_{\rm b}\geq 3$, we do not know if this feature is generic or restricted to small circuits. 

Part of the challenge in achieving further improvements for the TFIM is the extremely rapid convergence of variational energy with bond-dimension to the exact ground-state energy. We note that, in an actual quantum computation, resolving very small energy differences will become impractical due to large statistical sampling overhead. This issue is especially pronounced in the TFIM, likely due to its integrability and small central charge, $c=\frac12$ (the smallest of any minimal model). In particular, assuming a near-optimal variational ansatz, the effective correlation length scales with the number of bond qubits like $\xi \sim n_{\rm b}^{(3/c)\log 2}$ (at criticality), so that smaller $c$ yields larger correlation length, and more rapid convergence of finite-range correlations with the addition of bond qubits. 


\begin{figure}[tt]
\begin{center}
\includegraphics[width=0.95\columnwidth]{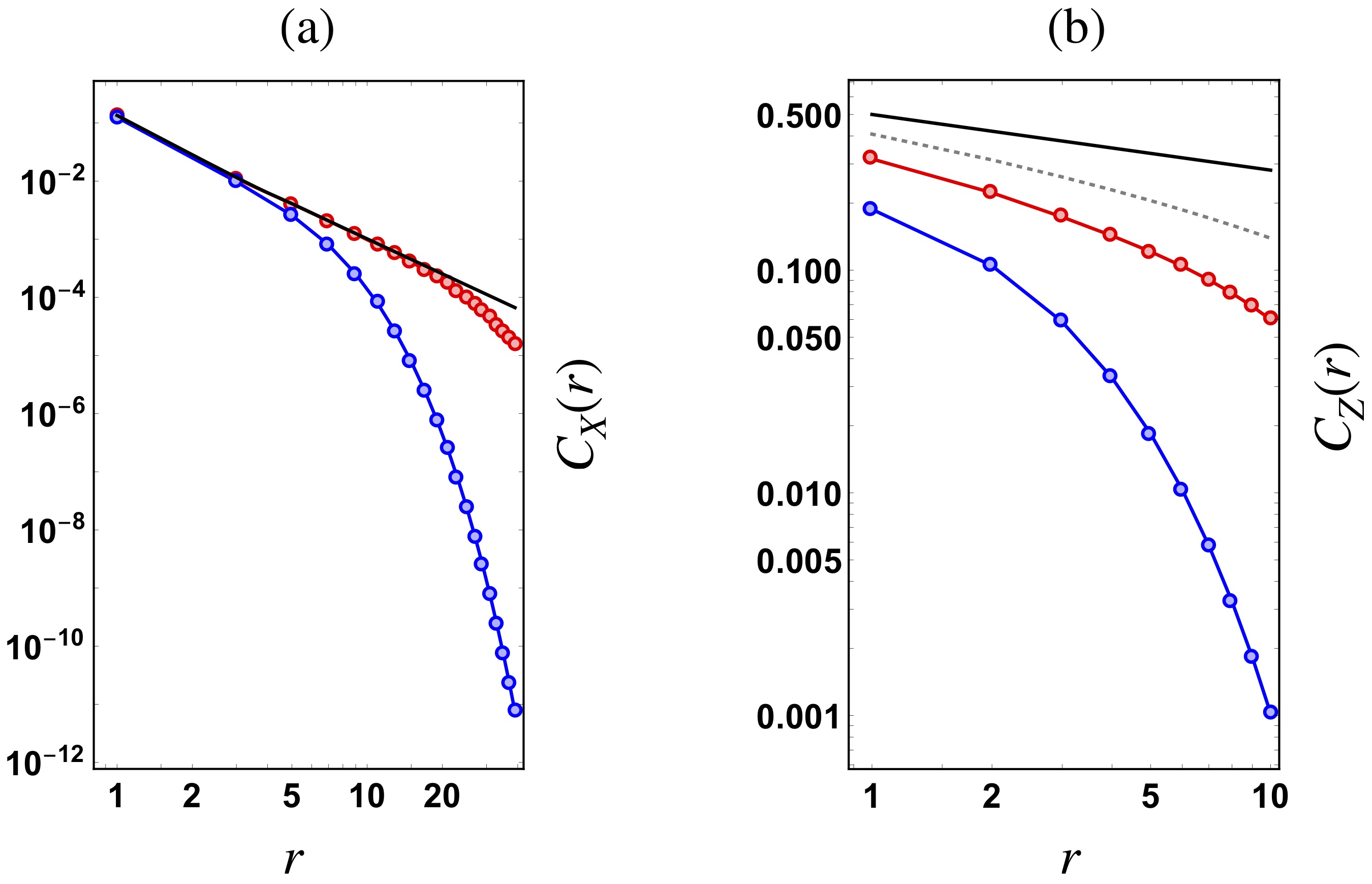}
\caption{Correlation functions of the transverse (a) and longitudinal (b) spin directions in the 1D TFIM.  Black lines are exact from fermionization, while the points are from holoVQE (blue points for $n_{\rm b}$=1 and red points for $n_{\rm b}=2$).  In (b) the $\chi=8$ curve is shown as well (black dashed line), though this has been obtained from direct MPS optimization as the numerical minimization for an $n_{\rm b}=3$ star-circuit was inconclusive.}
\label{fig:TFIM_correlations}
\end{center}
\end{figure}

Since the Hamiltonian is comprised of nearest-neighbor interactions, the holoVQE procedure only requires measurement of nearest-neighbor correlation functions.  Once the circuit is optimized one can freely use the holographic state preparation subroutine to extract any desired observables in the variational solution.  For example, in \fref{fig:TFIM_correlations} we show the critical ($J=h$) transverse and longitudinal connected spin-spin correlation functions: 
\begin{align}
	C_{X}(r) &=\<X_jX_{j+r}\>-\<X_j\>^2, \nonumber\\
	C_{Z}(r) &= \<Z_jZ_{j+r}\>-\<Z_{j}\>^2, 
\end{align}
where we've assumed translational invariance.  The points are obtained from the holoVQE method, while the solid black lines are exact results from fermionization. The holoVQE results show clear signs of the universal scaling behavior for the Ising transition, $C_{X}(r) \sim r^{-1}$, and $C_{Z}(r)\sim r^{-1/4}$ over moderate length scales (as much as 20 sites for the transverse correlations), despite using no more than three qubits.

%
%

%
%
%

\section{Dynamics}
Calculating the dynamical properties of interacting quantum systems is an essential challenge for practical applications such as predicting chemical kinetics, computing non-equilibrium electronic and optical properties of quantum materials and devices, and analyzing NMR spectra~\cite{altman2019quantum}. Quantum dynamics also underpins foundational scientific questions ranging from the nature of thermalization and quantum chaos, to properties of quark-gluon plasmas in heavy-ion collisions, to understanding cosmological scenarios for defect production. Despite its fundamental importance, simulating quantum dynamics remains among the most challenging tasks for classical computers,  generically requiring exponential classical resources even in low-dimensional systems where ground-states can be efficiently simulated. This area is therefore a promising candidate for achieving a practical quantum advantage on near-term quantum hardware.

In the following, we develop a quantum simulation algorithm that incorporates the holographic representation of MPS initial states described above to simulate time-evolution of an initial state under a quantum quench:
\begin{align}
|\psi(t)\> = \mathcal{T}\{e^{-i\int_0^t H(s) ds}\}|\psi(0)\>.
\end{align}
Here, the initial state $|\psi(0)\>$ is represented by an MPS (potentially with interesting correlations and entanglement), and $H(t)=\sum_\alpha h_\alpha$ is any geometrically-local time-dependent Hamiltonian, where each term $h_\alpha$ acts on at most $k$ adjacent physical sites. We dub this technique holographic quantum dynamics simulation (holoQUADS).  HoloQUADS enables a simulation of arbitrary time-ordered correlation functions using only $\sim {\rm poly}(t)\log \mathcal{Q}+\log(\chi)$ qubits, which is {\it independent} of $\ell$ \footnote{The exact polynomial dependence on the evolution time $t$, as well as the dependence on $k$, depends on the chosen Hamiltonian simulation algorithm}. By comparison, the classical resources required to simulate time-evolution from MPS initial states generically scale exponentially with $t$ due to rapid entanglement growth, even in $1$D systems~\footnote{A few interesting but non-generic exceptions to this rule include disordered, glassy dynamics of many-body localized systems, or dynamics very close to the ground state such as in a quench under a local perturbation.}.

\subsection{Holographic quantum dynamics simulation (holoQUADS)}
If the dynamics we care about is naturally generated by some circuit of depth $r$, we can proceed immediately with a holographic circuit construction as detailed below.  If we are concerned with continuous time evolution under a Hamiltonian $H(t)$, the first step is to approximate the resulting unitary by a circuit consisting of $r$ layers. There are many ways this approximation can be accomplished (see Refs.\ \cite{Haah_2018,childs2019theory} for a helpful review of some state-of-the-art Hamiltonian simulation techniques), with different techniques having different scalings of $r$ with time, qubit number, and desired simulation accuracy.  For our purposes, it suffices to note that algorithms exist for which, at fixed error, $r$ scales no worse than $\sim \ell ^{\varepsilon} t^{1+\varepsilon}$ for any $\varepsilon>0$ \cite{childs2019nearly, Haah_2018}. We consider this scaling to be essentially linear in $t$ and {\it independent} of $\ell$.  Taking a more pragmatic approach, we note that simple product formulas based on Trotterization generally require a number of layers that is far smaller than most rigorous bounds suggest, and---at least for local observables like 2-body correlation functions---produce accurate results using a modest (and system-size independent) number of time steps.


\begin{figure}[t]
\begin{center}
\includegraphics[width=0.95\columnwidth]{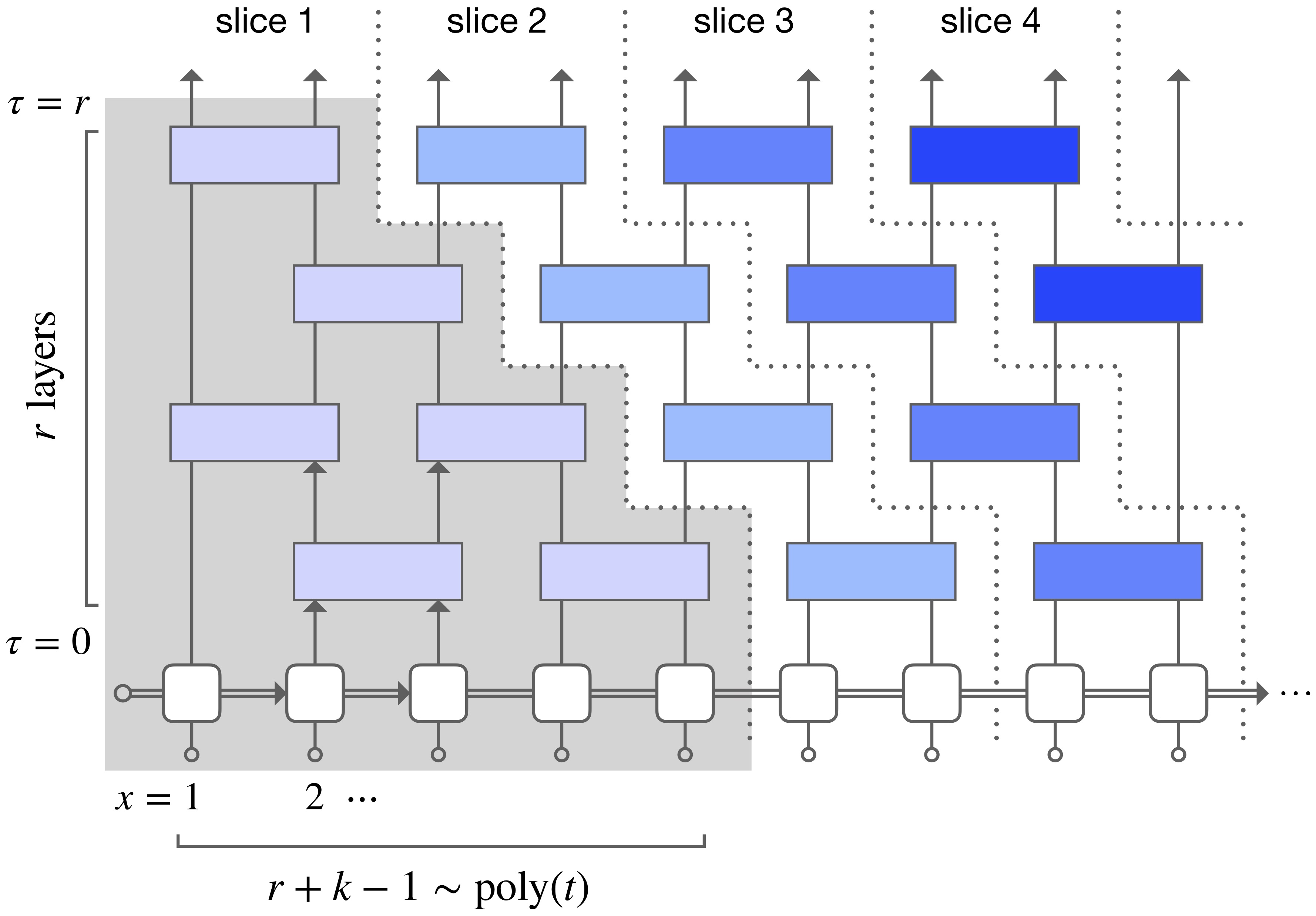}
\caption{Holographic time evolution of a matrix product state for nearest-neighbor interactions ($k=2$). The circuit can be evaluated by first executing all gates in the past causal cone (gray shaded region) of the qubits exiting the top left corner of the circuit.  The remaining slices can be executed in order by: (a) resetting the qubits exiting the top of the previous slice, (b) using the reset qubits to extend the MPS in space, and (c) applying all time-evolution gates in the current slice.}
\label{fig:time_evolution_circuit}
\end{center}
\end{figure}

Next, we seek a holographic description of the state resulting from applying this discretized evolution to the initial MPS. To do so, it is useful to adopt space-time-inspired terminology to discuss different geometrical regions of the circuit. Denote the layer of the circuit by a discrete time index $\tau$, and the position along the spin chain by $x$. Each wire in the circuit has an implied directionality, as shown by the arrows in \fref{fig:time_evolution_circuit}.  We define the past light cone of the point $(x,\tau)$ to be the set of all points $(x',\tau')$ from which one can arrive at $(x,\tau)$ by flowing along the circuit in a forward direction, exiting gates along any outgoing wire.
Inspection of \fref{fig:time_evolution_circuit} shows that measurements on the first $k$ ($k=2$ in this example) sites of the chain depend on the past light cone of the $k^{\rm th}$ qubit (gray-shaded-region of  \fref{fig:time_evolution_circuit}). The circuit in this region can be implemented using only the physical qubits for the first $r+k-1$ sites  (with $r$ scaling polynomially in $t$ for fixed error) along with the $\log_2(\chi)$ bond qubits: First implement the unitary circuits to prepare the MPS state within the gray-shaded region at $\tau=0$, and then apply the layers of gates from $\tau=0$ to $\tau=r$ that fall within this region. The first $k$ physical qubits at $\tau =r$ can be measured in any desired basis, and then reset and reused to represent the next $k$ physical qubits at $\tau=0$. These can be initialized into the correct state for the initial MPS by a horizontal (left-to-right) sequence of interactions with the bond qubits, and then propagated diagonally to $\tau=r$ by acting with the remaining gates lying within the past causal cone of the $2k^{\rm th}$ qubit at $\tau=r$, labeled as the diagonal ``slice 2'' in \fref{fig:time_evolution_circuit}.

\begin{figure}[t]
\begin{center}
\includegraphics[width=0.4\columnwidth]{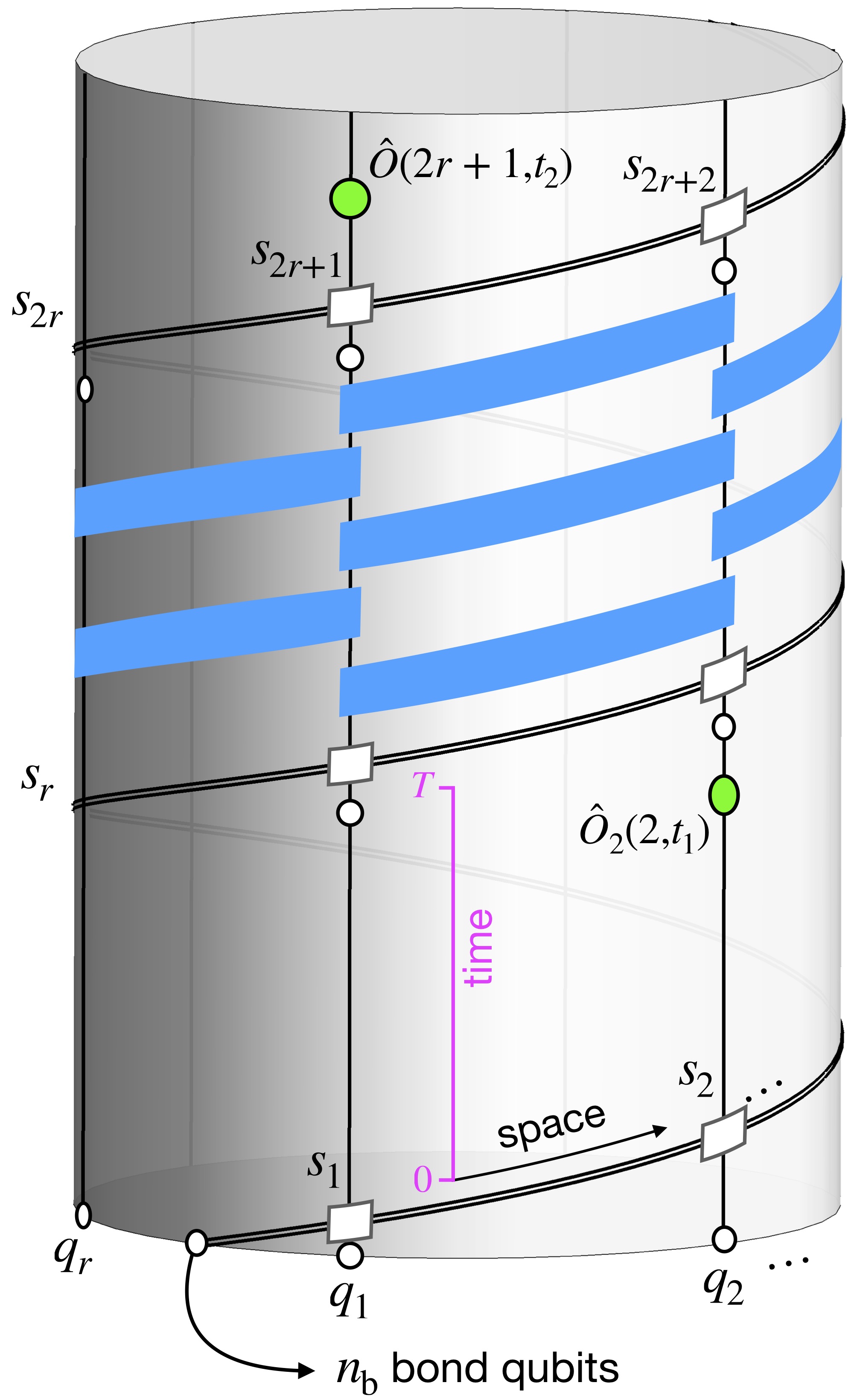}
\caption{Circuit for holographic time evolution of an MPS, obtained from \fref{fig:time_evolution_circuit} by attaching wires at exiting the top of the circuit to those at the bottom wherever the respective exiting qubits are reset and reused at the bottom of the circuit. The space direction wraps diagonally around the cylinder indefinitely (the site of the physical system at position $x$ is labeled $s_{x}$), and the circumference of the cylinder, which determines the required number of qubits, is determined by the evolution time. A time-ordered correlation function $\langle\hat{O}(x_1,t_1)\hat{O}(x_2,t_2)\rangle$ is obtained by measuring the corresponding operators at appropriate places in the circuit, as shown here (green circles representing measurements) for the example $x_1=2$, $x_2=2r+1$.}
\label{fig:time_evolution_cylinder}
\end{center}
\end{figure}

Repeating this process, one implements the full time-evolution circuit from left to right by sequentially implementing left-facing diagonal slices of the circuit. Effectively the top of the circuit is being sheared and reattached to the bottom, such that the actual circuit fits naturally on the geometry of a cylinder as drawn in \fref{fig:time_evolution_cylinder}.  By measuring the physical qubits at desired space-time points, one can reconstruct the time-ordered  correlation functions of any local operators (relevant for dynamical response both near and far from equilibrium). Further, out-of-time ordered correlators that provide insight into thermalization, scrambling, and many-body quantum chaos can be simulated by including intervals of reverse-time evolution.

One can alternatively view this construction as a generalization of the holographic-MPS representation obtained by slicing the time-evolved-MPS circuit into diagonal slices whose boundaries are left-future-null-trajectories, and considering all qubit lines entering the slice from the left as ``bond-qubits", and those exiting the slice vertically as physical qubits. With this interpretation, one can measure entanglement Renyi entropies of the physical chain by measuring the corresponding quantities of the bond-qubits as described above for general holographic MPS.

\subsection{Comparison to classical methods}
We now contrast this method with classical time-evolution techniques for simulating quantum dynamics. In $1$D systems the leading classical method for simulating time-evolution of an MPS is time-evolving block decimation (TEBD). TEBD works by converting an infinitesimal time evolution step $e^{-iH\Delta t}\approx (1-iH\Delta t)$ into a matrix-product operator (MPO), applying the MPO to the initial MPS state, and reinterpreting the result as an MPS with bond space given by the tensor product of the MPO and MPS bond spaces.

At each stage, the tensors of the MPS are compressed (if possible) by discarding subleading singular values below the target accuracy. The compression step is effective if little entanglement is generated during the Trotter-step compared to the maximum possible, but does not save classical resources in cases where significant entanglement is generated during each Trotter-step, e.g. as in a strong quench with a thermalizing Hamiltonian. In fact, in extreme examples with maximally entangling dynamics, such as simulating stroboscopic dynamics of random circuits~\cite{nahum2017quantum,von2018operator}, no compression whatsoever can be achieved. In contrast, holoQUADS exhibits polynomial scaling of the required qubit resources with evolution time regardless of the amount of entanglement generated per Trotter step, and will exhibit a maximal advantage in cases where low-rank classical compression is ineffective, or in higher dimensions where bond dimension can be prohibitively high for classical simulation from the outset. 

As an aside, we note that applying an MPO to an MPS (which forms the basis of many classical methods) cannot be directly implemented holographically by unitary circuits plus measurement, since application of an MPO does not generally preserve the right canonical form of an MPS. We leave as an open question for future work whether holoQUADS can be generalized to incorporate different, non-MPO-based quantum analogs of such compression schemes that operate efficiently on exponentially large bond spaces~\cite{lloyd2014quantum,rebentrost2018quantum,romero2017quantum} to further save on qubit resources.



\section{Discussion}
These holographic methods will provide a quantum advantage for situations where classical MPS techniques are intractable due to prohibitively high bond dimension. Physically relevant examples include ground-state preparation of systems in dimensions $D>1$, and time-evolution under thermalizing Hamiltonians. State-of-the-art classical techniques run out of steam for $2{\rm D}$ spin-systems of widths of around $10$ spins (less for gapless systems), or 1D time evolution with thermalizing dynamics over a few tens of interaction times. Holographic quantum algorithms might provide quantum advantage on these tasks with as few as 30-40 qubits, which are employed to directly tackle the difficult, highly-entangled quantum aspects of these problems, rather than spending these resources to capture large sections of Hilbert space that are not accessed in physically relevant systems.

For higher-dimensional simulations, the classical simulation cost for MPS techniques grows exponentially in the system width, even for area-law entangled states. One could employ a holoVQE method analogous to higher dimensional DMRG, obtained by slicing the system into a 1D stack of ($D-1$)-dimensional cross-sectional slices, and specifying a circuit architecture to implement a quantum channel connecting one slice to another. The exponentially large bond dimension of the resulting MPS could be captured with polynomially many bond qubits. Importantly, for physically relevant low-energy states, MPS provide an exponential compression over a full wave-function description even in $D>1$, so that the holographic representation affords substantial savings in qubit resources.

A more intrinsically higher-D generalization of these holographic methods would be to implement holographic simulation of isometric tensor networks~\cite{zaletel2020isometric}. These recently constructed higher-D generalizations of MPS capture a broad range of correlated states including (non-chiral) long-range entangled topological orders~\cite{soejima2020isometric}, and can be implemented straightforwardly as unitary circuits between physical qubits (now for each site in the lattice rather than each cross-sectional slice) and ancillary bond qubits. The isometric tensor networks invoke an explicit ordering of operations, and can be recast as a quantum channel that can be implemented holographically by resetting and reusing physical qubits that have already completed their participation in the circuit. The advantage of this technique over the usual boustrophedonic sweeping for 2D DMRG techniques is that it imposes a natural geometrically local 2D structure onto the physical and bond qubits.

It would also be desirable to extend these techniques to treat fermionic systems, using fermionic MPS representations~\cite{bultinck2017fermionic}, in order to simulate electronic materials. Other topics for future study include identifying effective heuristics for iteratively increasing bond-dimension to improve the holoVQE accuracy, and developing a formal and systematic understanding to the class of states that can be efficiently represented holographically using large qubit numbers, but reasonable circuit depths, to implement each MPS tensor.

\noindent\textit{Acknowledgements -- } We thank Romain Vasseur, Aaron Friedman, and Michael Zalatel for helpful discussions.

\end{document}